\documentclass[twocolumn,prb,showpacs]{revtex4}

\usepackage{amssymb,amsfonts,amsmath,graphicx,bm}
\begin{document}

\title{Surface electron band structure and VLEED reflectivity for Al(111)}
%\date{}
\author{M. N. Read}

\affiliation{School of Physics, University of New South Wales, \\ 
Sydney NSW
2052, Australia}
\email{mnr@phys.unsw.edu.au}

\begin{abstract}
The 2D layer Green function scattering method is used to calculate the energy of surface states and resonances at $\overline{\Gamma}$ for Al(111) for both below and above the vacuum level. The surface barrier potential is represented by an empirical form.  The above vacuum level surface electron band structure for this surface has not been calculated before and it is important in understanding many surface phenomena. The geometric structure of the Al(111) surface is known from intensity analysis in low-energy electron diffraction at energies 60 -- 450 eV. The details of the surface barrier for Al(111) were obtained from a match with the below vacuum level experimental energy position of the first Rydberg surface resonance and the Shockley surface state at $\bm{k}_{\parallel} = 0$ ($\overline{\Gamma}$). The calculation was then extended to the above vacuum level case for 0 -- 27 eV with the inclusion of inelastic electron interactions. Tamm-type resonances at 6.9 eV and possibly also at 8.3 eV, a Shockley-type resonance at $14.0\pm0.5$ eV and a series of Rydberg (image) resonances near 24 eV all above vacuum level are found at $\bm{k}_{\parallel} = 0$. The same 2D layer Green function scattering method using the same input data was then used to calculate the intensity of the 00 beam for $\bm{k}_{\parallel} = 0$ (normal incidence) in very low energy electron diffraction (VLEED) from this surface in the energy range 0 -- 65 eV. Features in the VLEED intensities are found due to the Shockley and Rydberg resonances. Experimental data from over 26 years ago found surface features near the energies found in this work. Beam intensities from low energy electron microscope (LEEM) measurements at normal incidence and new data from other surface spectroscopies could provide experimental confirmation of the resonances predicted in this work.
\end{abstract}
\maketitle

\section{Introduction and Aim}

In this work we begin an examination of the unoccupied surface electronic band structure for Al(111) in the energy range 0 -- 27 eV above the vacuum energy level.  Surface-state resonances for these energies are important in understanding surface properties and in the analysis of many surface spectroscopies including e.g. photoemission and inverse photoemission. At present there does not appear to be any other calculation of these bands for this crystal surface. We will calculate energy states and resonances both below and above vacuum level using the layer-by-layer scattering approach. The relationship between surface and bulk electron band structure and low-energy electron diffraction (LEED) reflectivities (or intensities) has been demonstrated previously. \cite{jennings1} Such a LEED analysis provides one means of verifying the accuracy of the calculated above vacuum level surface band structure. Hence we will also calculate very low energy reflectivities in this energy range to compare with experimental data. These reflectivities are calculated by the same layer-by-layer scattering method and use the same input information for the crystal surface potentials as the band structure calculation. 

For the higher energy range of 60 -- 450 eV above vacuum level successful LEED analyses have been performed for Al(111)\cite{jepsen2,jona3,neilsen4,noonan5,burch6} and these also reflect the electron energy bands at these energies. Most of these analyses have been at normal incidence for the non-specular beams only since the reflected 00 beam cannot be measured directly in the usual LEED set-up. Geometric structure and electronic properties have been extracted from these analyses for this energy range. Similarly, surface and bulk energy bands are known for the below vacuum level energy region chiefly from experimental photoemission and inverse photoemission spectroscopies. For the connecting energy range 0 -- 65 eV above vacuum level, the surface band structure is essentially unknown. This energy range is complicated by the significant variation of the complex electron self-energy and also details of the crystal bulk and surface scattering potentials including the surface barrier. By utilizing the known structural and non-structural properties determined from experiment for the adjacent energy ranges we will predict properties for the intermediate energy range that can be verified by beam intensity analysis in very low-energy electron diffraction (VLEED) and in low energy electron microscopy (LEEM). New experimental data is now possible because the LEEM apparatus can measure the specular 00 beam reflectivity at normal incidence and very low energies.

\section{Method}  

For the calculation of the surface electronic band structure and VLEED reflectivity the same scattering approach is applied. This is the 2D layer Korringa-Kohn-Rostocker (KKR) Green function scattering method of Kambe \cite{kambe7} and transfer matrix method of McRae \cite{mcrae8} for combining 2D scattering layers. The surface barrier potential is specified by an empirical form. Because of the increased sensitivity to bulk scattering potentials, $U(r)$, two potentials that have been used to calculate below vacuum level bulk-band structures are used. The incident electron self-energy, $\Sigma(E,\bm{k})$, is given by
\begin{equation}
\Sigma(E, \bm{k}) = U_0(E, \bm{k}) - i U_{\text{in}}(E, \bm{k})		                
\end{equation}
where  $U_0(E, \bm{k})$  is the crystal inner potential and $ U_{\text{in}}(E, \bm{k})$ is the inelastic scattering potential. \cite{pendry9} Both are known to vary significantly with energy $E$ (and possibly also momentum $\bm{k}$) in the energy range above the vacuum level. The lattice constant for f.c.c. Al was taken as 4.0496 \AA\ for room temperature of 300 K. \cite{coleridge10} Recent LEED determinations of surface structure have found an expansion of the surface atomic layer of $\sim1.4$ \% and smaller variations from bulk value for subsurface layers \cite{burch6} but these are not significant for the present study. Similarly electron-phonon interactions at 300 K are not included as yet. 

For the case of zero inelastic scattering $(U_{\text{in}}(E, \bm{k}) = 0)$, values of energy, $E$, and crystal momentum parallel to the surface, $\bm{k}_{\parallel}$, for which 
\begin{equation}
\text{det} \; [\bm{M}'] = 1		                
\end{equation}
correspond to total reflection from the crystal substrate and hence a surface-projected bulk-band gap. \cite{read11} $\bm{M}'$ is the scattering matrix containing amplitude reflection coefficients for propagating plane waves from the semi-infinite crystal substrate. \cite{mcrae8} Surface states and resonances are located for any $ U_{\text{in}}(E, \bm{k})$ by determining values of $ (E, \bm{k}_{||})$ for which
\begin{equation}
\text{det} \; [\bm{I} - \bm{S}^{\text{\bf{II}}}\bm{M}] \;\;\;\;\; \text{is a minimum. \cite{mcrae12}}		                
\end{equation}          
$\bm{M}$ is the full semi-infinite crystal scattering matrix (propagating and evanescent plane waves) and $\bm{S}$ is the surface barrier potential scattering matrix. \cite{mcrae13} The sub-matrix $\bm{S}^{\text{\bf{II}}}$ gives amplitude reflection coefficients describing internal scattering at the surface barrier potential from inside the crystal surface. The above condition determines at which energies the amplitude of the wavefunction passes through a maximum value in the surface region and corresponds to the electron being permanently or temporarily trapped in a surface state or resonance.

Reflectivities for VLEED beams are calculated by the method of McRae \cite{mcrae13} using the same $\bm{M}$ and $\bm{S}$ matrices. 

\section{Calculated surface band structure for Al(111) at $\overline{\Gamma}$}

\begin{figure}[b]
\includegraphics[scale=0.55]{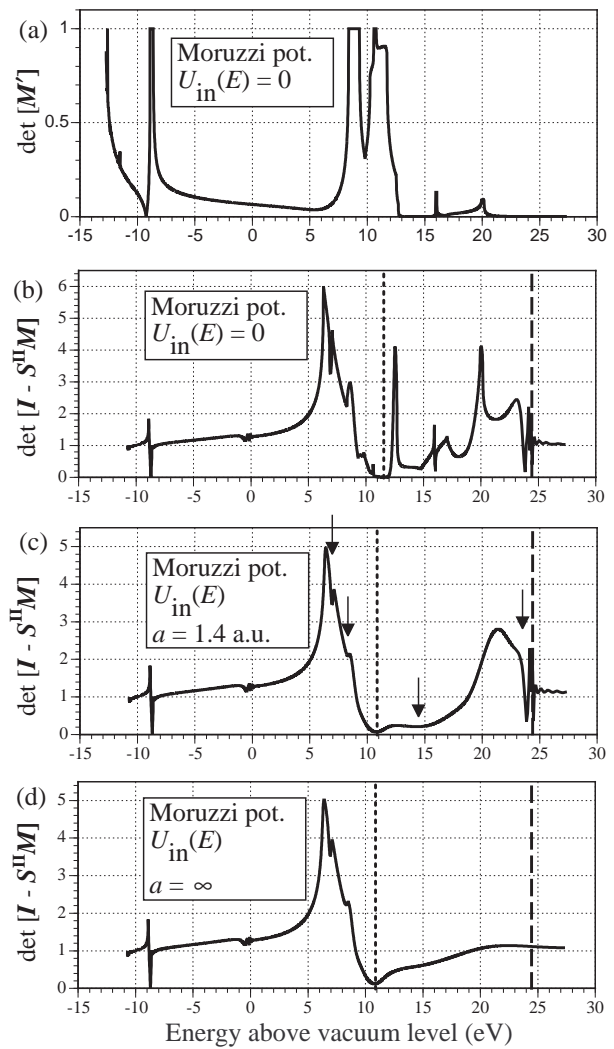}
\caption{Calculations in the present work for surface-projected bulk-band gaps from det $[\bm{M'}]$ and energy positions of surface states and resonances from det $[\bm{I} - \bm{S}^{\text{\bf{II}}} \bm{M}]$ for Al(111) and $\bm{k}_{\parallel} = 0$ ($\overline{\Gamma}$ point) using the Moruzzi {\it et al.} bulk potential and surface barrier. For frames (c) and (d), the values of $U_{\text{in}}(E)$ are shown in Figs.~4 \& 5. The termination of $U_{\text{in}}(E)$ at the surface is specified by a Gaussian function with half-width $a$. The downward arrows in frame (c) indicate the energy of surface resonances above the vacuum level. The two vertical lines indicate the energies at which the \{10\} \& \{01\} sets of six plane waves become propagating in the crystal (dotted line) and in the vacuum at 24.49 eV (dashed line).} 
\label{fig1}
\end{figure}

Fig.~1(a) shows the calculation of the band gaps of Al(111) for  $\bm{k}_{\parallel} = 0$ ($\overline{\Gamma}$ point) using the Moruzzi {\it et al.} self-consistent bulk potential \cite{moruzzi14} where exchange is given by a local-density functional. Fig.~2(a) is the same as Fig.~1(a) except that the Snow self-consistent bulk potential \cite{snow15} is used where exchange is given by the Slater approximation. In these calculations 7 plane waves were included for interlayer scattering and 6 phase shifts for intralayer atomic scattering for energies up to 27 eV. The 7 plane waves have surface reciprocal-net components 00, the triply degenerate set \{10\} and the triply degenerate set \{01\} for the surface net with primitive vectors at $120^{\circ}$. An energy interval of 0.01 eV was used for these calculations. The surface-projected bulk band gaps for $\bm{k}_{\parallel} = 0$ ($\overline{\Gamma}$ point) and no inelastic scattering $ (U_{\text{in}}(E, \bm{k})=0)$ are given by Eq.~(2) and are projections of bulk bands for $\Lambda_1$ symmetry for the $\Gamma(\Lambda)L$ direction.  Connelly \cite{connelly16} has extended the bulk band calculation for the Snow potential to above vacuum level energies but only for $\Gamma(\Delta)X$. A calculation by Szmulowicz and Segall \cite{szm17} that gives essentially the same result as Connelly's result for $\Gamma(\Delta)X$, also shows  $\Gamma(\Lambda)L$. Using the bulk band structure in Fig.~1 of Ref.~17 we see that the band gaps near $-9$ eV in Figs.~1 (a) and 2 (a) are due to the bulk gap bounded by the $L_1$ and$L_{2'}$ points (all energies are referred to the vacuum level unless specified otherwise). There does not appear to be experimental values for these energy points \cite{levinson18} and the Snow potential predicts a smaller gap-width than the Moruzzi {\it et al.} potential. Similarly the two higher energy gaps in the vicinity of 10 eV have different energy widths and position. From Ref.~17 we see that these two surface-projected bulk gaps arise from gaps bounded by $\Gamma_{25'}$ and $\Gamma_{15}$ points and by $L^u_1$ and $L^u_{2'}$ points respectively. For the energy range $0 - 65$ eV the band structure is complicated because of empty $3d$ bands at $\sim15$ -- 20 eV and $4f$ bands at $\sim30$ eV. \cite{levinson18}

\begin{figure}[t]
\includegraphics[scale=0.55]{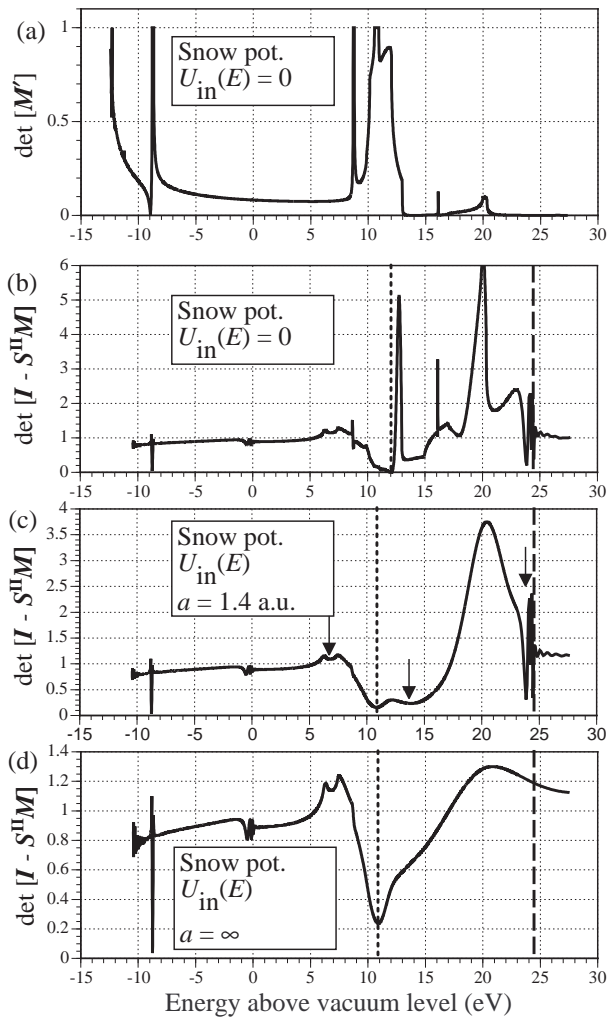}
\caption{Same as Fig. 1 except using the Snow bulk potential and surface barrier.}
\label{fig2}
\end{figure}

For the surface band calculation we represent the crystal-vacuum interface with an empirical potential barrier $U(z)$ with image tail $1/z$ into the vacuum where $z$ is the perpendicular distance from the surface plane. The origin $z = 0$ is at the centre of the top row of atoms and the crystal is located at +ve  $z$  values.  The Fermi energy with respect to the muffin-tin zero of potential, $E_{\text{f}}$, is 8.5 eV and 8.2 eV for the Moruzzi {\it et al.} and Snow potentials respectively and the experimental work function, $\phi$, is $4.24 \pm  0.02$ eV. \cite{grepstad19} This gives the inner potential or barrier height up to the vacuum level as $U_0(E = 0) = E_{\text{f}}  + \phi = 8.5 + 4.24 = 12.74$ eV and 12.44 eV for Moruzzi {\it et al.} and Snow potentials respectively. The form of the surface potential is $1/(z - z_0)$ where $z_0$ is the image-plane position and this form joins smoothly to a cubic-polynomial type saturation at a point $z_1$ closer to the metal surface. This model is sufficiently close to the form found from {\it ab initio} non-local full potential density-functional calculations for the crystal-vacuum interface \cite{heinrich20} and has an advantage over other empirical models that the manner of the join to the crystal can be controlled. For this case we join the barrier to the bulk muffin-tin zero at $z = 0$. The model is described in detail by Malmstr\"{o}m and Rundgren. \cite{malm21} 

In order to determine the specific details of the barrier for this metal surface, we calculated surface states and resonances for $\overline{\Gamma}$ from Eq.~(3) for various choices of $z_0$ and $z_1$ and compared with experimental results. Below the vacuum level the inelastic potential is negligible and $U_{\text{in}}(E, \bm{k}) = 0$. At $\overline{\Gamma}$ a surface state has been detected by high-resolution angle-resolved photoemission spectroscopy (ARPES) at $4.56 \pm 0.04$ eV below $E_{\text{f}}$. \cite{kevin22} The first Rydberg image resonance has been detected at 3.75 eV above $E_{\text{f}}$ or $\sim0.5$ eV with respect to the vacuum level by $k$-resolved  inverse photoemission spectroscopy (KRIPES) \cite{heskett23,yang24} and by scanning tunnelling spectroscopy (STS). \cite{yang24} By variation of $z_0$ and $z_1$ we reproduced the energy position of the Rydberg resonance with $z_0$ at $-1.1$ a.u. and $z_1$ at $-2.0$ a.u. and $z_1$ at $-1.85$ a.u. for Moruzzi {\it et al}. and Snow potentials respectively. The lower energy surface state was then found to be at the experimental energy also for these $z_0$ and $z_1$ values. The resulting surface barrier $U(z)$ for the Moruzzi {\it et al.} potential is plotted in Fig.~3. The results of the calculations using Eq.~(3) and the above empirical surface barriers are shown in Figs.~1(b) \& 2(b) with minima at the correct energies below the vacuum level. The 00 plane wave emerges into the vacuum at energies above the barrier height at the vacuum level. Below the vacuum level at $\overline{\Gamma}$ only the 00 plane wave is propagating in the crystal and the other six plane waves are evanescent or attenuated in space. The surface band energies arise from normal incidence scattering of the propagating 00 plane wave between the crystal and barrier. This repeated scattering gives constructive interference and a maximum amplitude of wave function at these two energies as the energy increases from the muffin-tin zero to the top of the barrier. This corresponds to the electron being trapped in a surface state and temporarily trapped in a series of surface resonances of the crystal. These energy positions are particularly sensitive to the shape of the barrier because of the $\sim 8.3$ eV energy range spanned between them. The series of resonances arises from scattering near the top of the barrier where the phase change on scattering goes through cycles of $2\pi$ because of the long-range $1/z$ tail. The surface state near $-9$ eV in a surface-projected bulk-band gap of the crystal is of the Shockley type because of its strong energy dependence on the form of the barrier.  Because of this dependence surface states/resonances of this type are often described as arising from a symmetrical termination of the crystal. The wavefunction maximum lies close to the top row of atoms. The Rydberg resonances are of the same type but are distinguished by their strong dependence on the image tail of the surface barrier. In this case the wavefunction maximum lies well into the vacuum region.
      
\begin{figure}[t]
\includegraphics[scale=0.5]{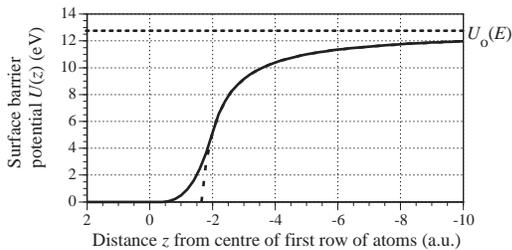}
\caption{Surface barrier potential (full line) for Al(111) and the Moruzzi {\it et al.} bulk potential. The image-plane position is at $z = -1.1$ a.u. and saturation away from the truncated image potential (dashed line) is at $z = -2.0$ a.u.}
\label{fig3}
\end{figure}

Having determined the details of the surface barrier from measured energies of a surface state and Rydberg resonance below the vacuum level we can now use this information to calculate the above vacuum level surface bands. At present there does not appear to be any other calculation of these bands for this crystal surface. We use the same surface barrier potential as that used for the below vacuum level surface band structure in Figs.~1 and 2 with $U_0(E, \bm{k})$ in Eq.~(1) given by $U_0(E = 0)$, i.e. no change in  the barrier height of $U_0(E = 0)$. This value should decrease in energy over the present energy range but until experimental results are obtained we have not included this variation in the calculation as yet. Any other variations of the shape and image-plane position of the barrier are considered to be small for the current $(E, \bm{k})$ values. Figs.~1(b) \& 2(b) also show minima in the above vacuum level energy range indicating surface resonances. The \{10\} \& \{01\} sets of six plane waves become propagating in the crystal at 11.76 eV in Fig.~1(b) and 12.06 eV in Fig.~2(b) (vertical dotted line) and propagating in the vacuum at 24.49 eV (vertical dashed line). The minimum at the vertical dotted line indicates the energy of the transition from evanescent to propagating for the six \{10\} \& \{01\} plane waves in the crystal and is not of the same character as the other minima. In this case these crystal-emerged plane waves are travelling parallel to the surface and do not produce the surface bands of the crystal. 

Between 5 eV and the vertical dotted line in Fig.~1(b) and Fig.~2(b) where there are a number of minima, the above six plane waves are evanescent in the crystal. Their energy position does not depend on the barrier shape but a maximum amplitude of wavefunction exists above the top row of atoms at these energies. In this case the surface layer of atoms is different from the bulk layers in that there are no neighbouring atom layers on the vacuum side of the top atom layer. Such resonances may be termed Tamm-type since they arise from what may be considered the unsymmetrical termination of the crystal at the surface. Between the vertical dotted line and 24.49 eV the above six plane waves are propagating in the crystal, are incident at different angles on the surface barrier and interact. The internally reflected plane waves then become incident on the crystal and scatter from it in a similar way to the 00 plane wave previously considered in the below vacuum level case. From Eq.~(3) minima are found corresponding to surface resonances. Above the vacuum level all surface bands are resonances because scattering at the crystal can now redistribute electron flux into the vacuum-emerged 00 plane wave. The lower energy minima indicate Shockley-type resonances. In this energy range the six plane waves also give rise to a series of Rydberg resonances where scattering occurs near the top of the barrier just before their emergence into the vacuum. 

However the above vacuum level bulk and surface band calculations mentioned above and calculated in frames (a) \& (b) of Figs.~1 \& 2 are not realistic because of the neglect of broadening due to electron-electron inelastic scattering at these energies. Surface-projected bulk-band gaps now become pseudogaps. In the present method, inelastic scattering of this type is included by the term $U_{\text{in}}(E, \bm{k})$ of the electron self-energy in Eq.~(1). For these calculations we use an energy dependence only, $U_{\text{in}}(E)$, found from an analysis by McRae from photoemission experimental data \cite{mcrae25} and this result has been plotted without the logarithmic scale in Fig.~4. The rapid rise corresponds to the bulk-plasmon excitation threshold at 11.3 eV above the vacuum level. \cite{pillon26} The termination of the crystal inelastic scattering potential $U_{\text{in}}(E)$ to give the surface barrier inelastic scattering potential is specified by joining a Gaussian function of height $U_{\text{in}}(E)$ and half-width $a = 1.4$ a.u. to $U_{\text{in}}(E)$ and this is illustrated in Fig.~5 for the case of $U_{\text{in}}(E) = 4.1$ eV.  This half-width is within the expected value for the range of the absorption potential at the surface. \cite{mcrae27} Smaller values would lead to unrealistically strong surface resonances.

\begin{figure}[h]
\includegraphics[scale=0.38]{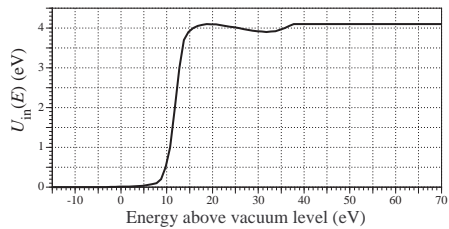}
\caption{Bulk inelastic scattering potential $U_{\text{in}}(E)$ for Al.}
\label{fig4}
\end{figure}

\begin{figure}[h]
\includegraphics[scale=0.5]{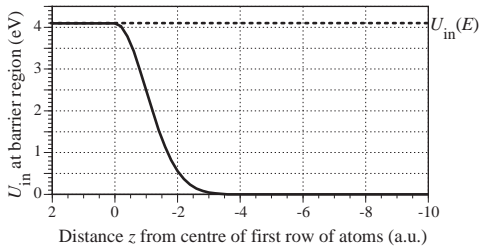}
\caption{Surface barrier inelastic scattering potential for Al(111) with $U_{\text{in}}(E) = 4.1$ eV and Gaussian function of half-width $a = 1.4$ a.u.}
\label{fig5}
\end{figure}

Figs.~1(c) \& 2(c) show the minima from Eq.~(3) when the above inelastic scattering potentials are included and indicate the expected surface bands that would be found experimentally from the present model. For $\overline{\Gamma}$ one Tamm-type surface resonance near 7 eV survives inelastic broadening and also another at 8.3 eV for the Moruzzi {\it et al}. potential. The strength of these resonances depends to some degree on the exact values of $U_{\text{in}}(E)$ in the energy range 5 -- 10 eV. One Shockley-type resonance survives at $\sim14.5$ and 13.5 eV for the Moruzzi {\it et al}. and Snow potentials respectively and is strong since the minimum from Eq.~(3) is $<0.3$. The Rydberg series of resonances is also strong. Although the energy dependence of the inner potential (also barrier height here) has not been included, the energy position of the Shockley resonance to be found from experiment should still be within a few eV of that found here. Also its occurrence does not depend on the detailed variation of $U_{\text{in}}(E)$ since this quantity has reached its maximum value by 15 eV.

In order to illustrate the effect of no internal scattering at the surface barrier the barrier absorption potential is extended far into the vacuum by letting the half-width $a$ approach $\infty$. Wavefunctions are damped out in the surface barrier region and the Shockley and Rydberg surface resonances do not exist as seen in Figs.~1(d) \& 2(d). The Tamm surface resonance is not affected since it is not associated with the surface barrier extension into the vacuum.

\section{Experimental detection of above vacuum level surface-state resonances} 

Experimental techniques that may detect features due to surface-state resonances above vacuum level energies include VLEED/LEEM, target  (or total) current spectroscopy (TCS),  surface soft-X-ray absorption spectroscopy (SSXA) and inverse photoemission spectroscopy (IPS).

We have calculated the LEED/LEEM 00 beam reflectivity (or intensity) for Al(111) at 300 K for normal incidence corresponding to $\bm{k}_{\parallel} = 0$ for 0 -- 27 eV to match the calculations for the surface band structure by using the same input data as in that calculation. We use the $\bm{M}$ and $\bm{S}$ scattering matrices in the layer-by-layer method previously described. The calculation was also extended from 27 to 65 eV in order to gain information about the value of the crystal inner potential $U_0$, for the higher energy range once experimental data is available. An energy interval of 0.1 eV was used with 19 beams and 8 phase shifts for energies from 27 -- 65 eV. The inelastic scattering potential is the same $U_{\text{in}}(E)$ used for the band structure calculation and is shown in Figs.~4 \& 5.

\begin{figure}[b]
\includegraphics[scale=0.5]{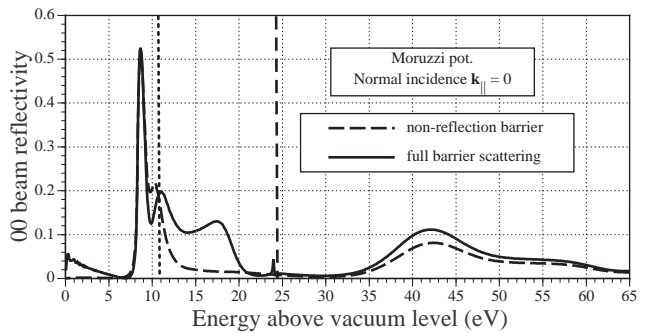}
\caption{Reflectivity (or intensity) of 00 beam on Al(111) at normal incidence using the Moruzzi {\it et al.} bulk potential with the same surface barrier as used in Fig.~1(c) and a non-reflection barrier. The two vertical lines indicate the emergence energies of the first six non-specular beams in the crystal (dotted line) and in the vacuum at 24.49 eV (dashed line).}
\label{fig6}
\end{figure}

Figs.~6 \& 7 show the calculated reflectivity using a non-reflection barrier for the Moruzzi {\it et al.} and Snow bulk potentials respectively. Here the bulk is terminated at a surface plane placed at the jellium discontinuity $z_j$ (half an interlayer spacing parallel to the surface). This step barrier of height $U_0$ at $z_j = - 2.21$ a.u. allows for transmission and refraction of all beams but no reflection. Hence any features that may depend on the detailed form of the crystal-vacuum interface are not produced. Figs.~6 \& 7 show that in this case the main features in the range 0 -- 27 eV are two peaks which arise from the surface-projected bulk-band pseudogaps corresponding to the gaps for zero inelastic scattering shown in Figs.~1(a) \& 2(a). The main difference between these two bulk potentials is the width of these gaps and hence the width of the peaks in the 00 reflectivity. 

\begin{figure}[t]
\includegraphics[scale=0.5]{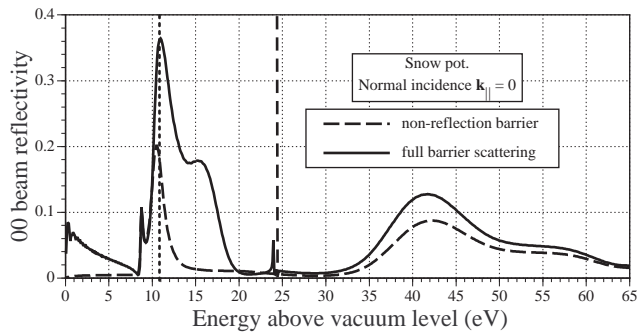}
\caption{Same as Fig. 6 except using the Snow bulk potential with the same surface barrier as used in Fig.~2(c) and a non-reflection barrier.}
\label{fig7}
\end{figure}

Also shown in these figures is the reflectivity for the case of full surface barrier scattering where the occurrence of surface resonances can be included in the calculation. Here both reflection and transmission are included for a surface barrier potential of the same form as that used earlier in the surface band calculation and shown in Fig.~3 for the Moruzzi {\it et al.} potential. Surface barrier scattering is specified in the $\bm{S}$ matrix. The $\bm{S}^{\text{\bf{II}}}$ sub-matrix that contains internal amplitude reflection coefficients for beams emerging from the crystal is the same as that used for the surface band structure calculation. However for the LEED case amplitude transmission coefficients for all beams passing through the barrier into the vacuum must also be specified and these are contained in the sub-matrix $\bm{S}^{\text{\bf{IV}}}$. In addition amplitude transmission and reflection coefficients for the incoming incident 00 beam in the vacuum also need to be specified and these are contained in the $\bm{S}^{\text{\bf{I}}}$ and $\bm{S}^{\text{\bf{III}}}$ sub-matrices. This incident 00 beam scattering has only a small effect near 0 eV where it is scattering just a few eV above the energy of the barrier height. Figs.~6 \& 7 indicate the emergence of the \{10\} \& \{01\} sets of six non-specular beams in the crystal (dotted vertical line) and in the vacuum at 24.49 eV (dashed vertical line).

The effect of surface resonances in the reflectivities of LEED beams has been explained by McRae. \cite{mcrae13} As for the band structure case and Shockley-type resonances their occurrence is detected in LEED where there is constructive interference in a sustained multiple scattering between surface barrier and crystal substrate for propagating pre-emergent beams. The energies where this type of scattering occurs corresponds to the electron being temporarily trapped in a surface resonance. When these non-specular beams become incident on the crystal in the LEED case some electron flux is redistributed to the vacuum-emerged 00 beam and other beams. This is an indirect process that occurs because of internal surface barrier reflection. There is also the direct process of flux into the vacuum-emerged beams by their single transmission through the surface barrier. Interference between these direct and indirect processes gives fluctuations in the reflectivity of vacuum emerged beams. As explained by McRae this can produce a peak, dip or combined peak-dip structure in the reflectivity profiles. This interference structure is centred at the energy of the surface resonance. For the Moruzzi {\it et al.} potential and barrier this occurs at 14.5 eV in Fig.~1(c) and in Fig.~6 we see a wide dip-peak structure centred near 14.5 eV and producing a pronounced peak at 17.5 eV.  Also a series of Rydberg resonance features occur near the vacuum-level threshold of the \{10\} \& \{01\} non-specular beams.  The first peak has a FWHM of $\sim 0.15$ eV and may be detectable in an experiment with suitable energy resolution. A similar strong dip-peak structure occurs for the Snow bulk potential and barrier at 15.5 eV as shown in Fig.~7. Here the centre of the LEED feature is at a lower energy corresponding to the value of 13.5 eV in Fig.~2(c) and the first Rydberg resonance feature has a similar energy width as that for the Moruzzi {\it et al.} potential and barrier. Thus either bulk potential gives a very strong surface resonance feature in the VLEED data at 16.5 $\pm$ 1 eV. This feature is wide in energy because the inelastic scattering potential, $U_{\text{in}}(E)$, has reached its maximum value of 4.1 eV and is of the Shockley type where the wavefunction maximum lies close to the crystal.

The Tamm-type resonance at $\sim 7$ eV in Figs.~1(c) \& 2(c) is stronger for the Snow potential and barrier and there is a possible indication of its presence in the very weak dip-peak structure in the VLEED 00 beam reflectivity near this energy in Fig.~7.

Some TCS experimental data \cite{jaklevic28} is available that shows the target current $I_c$ and its second derivative $d^2I_c/dE^2$. This data indicates the positions of peaks in the elastic reflection data at $\sim 13.7$, 16.0 and 21.5 eV with respect to $E_{\text{f}}$ and Rydberg peaks at higher energies. For comparison with Figs.~6 \& 7, the main three peaks are at 9.6, 11.8 and 17.3 eV with respect to the vacuum level and these results tend to confirm the occurrence of the Shockley surface barrier resonance feature near 16.5 $\pm$ 1 eV. The other two lower energy peaks correspond to the surface-projected bulk band pseudogaps. SSXA experimental data \cite{bachrach29} found a surface resonance at 12.1 eV with respect to $E_{\text{f}}$ or 7.9 eV with respect to the vacuum level and this is consistent with the Tamm-type resonances at $\sim 7$ eV in Figs.~1 \& 2 and at 8.3 eV in Fig.~1. There is no experimental VLEED data for 0 -- 27 eV for this surface of Al although there is early data for the (100) surface that also indicates strong surface resonance features. \cite{henrich30}

\section{Conclusion}

A Tamm surface resonance at 6.9 eV and also possibly one at 8.3 eV, a Shockley surface resonance at $14.0\pm0.5$ eV and Rydberg (image) resonances near 24 eV are predicted in the above vacuum level surface energy band structure at 300 K of Al(111) at $\overline{\Gamma}$ for 0 -- 27 eV. Strong features occur in this energy range in the 00 beam normal incidence VLEED reflectivity data due to the Shockley and Rydberg resonances. The present surface band structure calculations can be extended to other values of $\bm{k}_{\parallel}$ and the VLEED reflectivities to other incidence angles to examine the complete surface band structure. Before proceeding with these calculations and other refinements such as electron-phonon scattering and energy dependence of the self-energy it is most desirable to have confirmation of the major features present in the results predicted here for $\bm{k}_{\parallel}=0$. The LEEM experimental apparatus could measure this 00 spectra (and that of non-specular beams) for normal incidence at these low energies to confirm the occurrence of these surface resonances and possibly also distinguish between the two bulk Al potentials. Following this, further experimental data from VLEED, LEEM, TCS and other spectroscopies that includes other incidence angles would then be desirable to further analyse the above vacuum level surface band structure, surface potentials and inelastic processes for Al(111).

\end{document}